\documentclass{article}
\usepackage{FPSpro}
\newcommand{\qv}{{\vec q}}
\newcommand{\cO}{{\cal O}}
\newcommand{\half}{\frac{1}{2}}
\begin{document}
\title{ 
Using Lattice QCD and ChPT to obtain non-leptonic K-decay amplitudes
}
\author{
  Maarten Golterman\thanks{{}~~speaker at conference}        \\
  {\em Dept. of Physics, Washington University, St. Louis, MO 63130, USA} \\
  Elisabetta Pallante        \\
  {\em SISSA, Via Beirut 2--4, I-34013 Trieste, Italy}
  }
\maketitle
\baselineskip=11.6pt
\begin{abstract}

We summarize some of the difficulties that confront lattice 
calculations of non-leptonic kaon decay matrix elements.  We 
review some of the methods that have been proposed to overcome 
these difficulties, and discuss the importance of one-loop
ChPT in this respect, including the role of $O(p^4)$ operators.

\end{abstract}
\baselineskip=14pt
%
%
\vskip0.8cm
In this talk, we will discuss some aspects of
lattice computations of hadronic matrix elements, in particular
those relevant for non-leptonic kaon decays.  A more extensive review,
with additional references, can be found in ref.~\cite{mg}.  
In general, Lattice QCD provides us with euclidean correlation functions
which are typically computed at unphysically large values of quark masses 
and/or unphysical momenta (see below).  
In order to extract physical information, we can analytically
continue the unphysical lattice results using
chiral perturbation theory (ChPT), at least when we are 
concerned with the physics of Goldstone bosons of QCD.  

There are several methods
which are currently being used or investigated for 
the computation of kaon-decay rates.  The oldest
is to compute the $K\to\pi\pi$ matrix elements with not only the kaon
but also the pions at rest \cite{cbetal1}, and use ChPT in order to
continue to physical ({\it i.e.} energy-conserving) momenta.  Alternatively,
 $K\to\pi\pi$ amplitudes can also be related through soft-pion theorems to
the simpler $K\to\pi$ matrix elements which are, at least in principle,
easier to compute on the lattice \cite{cbetal2} (see also ref.~\cite{BPP})
Recent numerical
work employing both these methods can be found in refs.~\cite{num,gm}.
More recently, it has been suggested that  the finite size
of the spatial volume can be used as a tool to get to the physical
$K\to\pi\pi$ matrix elements \cite{ll}.  

In what follows we will always assume that we are dealing
with continuum-extrapolated lattice results, so that the usual ChPT
techniques directly apply. 
On the lattice, one extracts $K\to\pi\pi$ matrix elements from
euclidean correlation functions of the form
\begin{eqnarray}
C(t_2,t_1)\!\!\!\!&=&\!\!\!\!\langle 0|\pi(\qv,t_2)\pi(-\qv,t_2)\cO_{weak}(t_1)
K(0)|0\rangle \label{corr}\\
&\sim&\sum_n\langle 0|\pi(\qv,t_2)\pi(-\qv,t_2)|n\rangle\langle n|
\cO_{weak}|K\rangle\langle K|K|0\rangle\;e^{-E_n(t_2-t_1)-m_K t_1}\,,
\nonumber
\end{eqnarray}
for $t_1$ large. The dominant contribution to this correlation function 
at large euclidean times comes from the state with both pions at rest,
{\it i.e.} $\qv=0$ and $E_0=2m_\pi+O(1/L^3)$, while the desired physical state,
which has $|\qv|=\half\sqrt{m_K^2-4m_\pi^2}$, is buried
in the tower of excited states of eq.~(\ref{corr}) \cite{mt}.  Note that  for 
$|n\rangle=|\pi(\qv=0)\pi(\qv=0)\rangle$  
energy is injected by the weak operator.

These simple observations lead us to the three aforementioned methods.
In the approach proposed in ref.~\cite{ll} one may work in a 
finite volume, in which the smallest non-zero
momentum $|\qv|=2\pi/L$ is chosen such that $m_K=E_1\sim
2\sqrt{m_\pi^2+\qv^2}$ (omitting interaction effects).  If one can single out
the first excited state, this gives the desired matrix element at
finite volume, to which a correction factor can be applied to
convert to infinite volume \cite{ll}.  At the physical values of meson masses,
this requires a rather large volume of order $m_\pi L\approx 4$, 
or $L\approx 6$\ fm.
It will be interesting to see whether this method
can be made to work in practice.  This proposal together with a related 
approach \cite{LMST} is further discussed
by Guido Martinelli in these proceedings \cite{gm}.

A second, technically much simpler, 
method is to compute the unphysical matrix element
$\langle \pi(\qv=0)\pi(\qv=0)|\cO_{weak}|K\rangle$, and use ChPT to
convert it to the physical one \cite{cbetal1}.  
This approach  overcomes the problems implied by the Maiani-Testa theorem 
\cite{mt} since in the unphysical configuration with both pions at rest no 
final-state interaction phase is generated. A third method
is based on the observation that ChPT relates $K\to\pi\pi$ matrix elements to
$K\to\pi$ and $K\to 0$ transitions \cite{cbetal2}.
Issues which arise in the use of the latter two methods are: 1) the size
of chiral corrections, and related, 2) the role of $O(p^4)$ ChPT operators,
3) finite-volume effects, and 4) quenching.  
Issues 3 and 4 can also be investigated through
the use of one-loop ChPT.

As an example, let us consider the $\Delta I=3/2$  decay
$K^+\to\pi^+\pi^0$.  To one loop the physical matrix element is given by
\begin{equation}
\frac{[K^+\to\pi^+\pi^0]_{phys}}{m_K^2-m_\pi^2}\propto
\alpha^{27}
\left[1+{\rm chiral\ logs}+d_K\frac{m_K^2}{(4\pi f)^2}+d_\pi\frac{m_\pi^2}{(4\pi f)^2}
\right]\ , \label{phys}
\end{equation}
while for the unphysical matrix element with both pions at rest and 
$M_K=M_\pi=M$
the degenerate meson mass on the lattice \cite{glp} one has
\begin{eqnarray}
\frac{[K^+\to\pi^+\pi^0]_{unphys}}{M^2}
&\propto&\alpha^{27}(N)
\left[1+{\rm chiral\ logs}+d_s\frac{N}{2}\frac{M^2+M_{sea}^2}{(4\pi f)^2}
\right.\nonumber\\
&&\left.+d_v\frac{M^2}{(4\pi f)^2}
+\frac{17.8}{M L}+\frac{12\pi^2}{(M L)^3}\right]\ .
\label{unphys}
\end{eqnarray}
$N$ is the number of (degenerate) dynamical fermions ($=$sea quarks) on the
lattice, with corresponding Goldstone-boson mass $M_{sea}$
(which on the lattice does not have to be equal to the valence-meson mass 
$M$).  The quenched
approximation corresponds to $N=0$ (see ref.~\cite{mg} 
for more details).  The low-energy constant (LEC) $\alpha^{27}(N)$ is that of
the real world for $N=3$, even if the quark masses on the lattice
are not at their physical values \cite{shsh}.  
In the fully quenched theory, it may have a different value.  
Similar considerations also apply to the linear
combinations of $O(p^4)$ LECs $d_{K,\pi,s,v}$.  In addition, the linear
combinations showing up in the unphysical matrix element can be different
from those of the physical one:  For arbitrary (on-shell) momenta $p_K$,
$p_{\pi_1}$ and $p_{\pi_2}$ the contribution from $O(p^4)$ operators looks like
\begin{equation}
({\rm tree\ level})\times
\frac{1}{(4\pi f)^2}\left(AM_K^2+BM_\pi^2+Cp_{\pi_1}\cdot p_{\pi_2}
+Dp_K\cdot(p_{\pi_1}+p_{\pi_2})\right)\ .\label{gen}
\end{equation}
In the physical case $p_K=p_{\pi_1}+p_{\pi_2}$ so that the expression in
parentheses becomes $(A-\half C-D)M_K^2+(B+C)M_\pi^2$, whereas in the
unphysical case $p_{\pi_1}=p_{\pi_2}=({\vec 0},iM_\pi)$ and 
$p_{K}=({\vec 0},iM_K)$  lead to
the combination $AM_K^2+(B-C)M_\pi^2-2DM_KM_\pi$, which does not yield the 
same information.
The upshot is that one can determine the leading-order LECs, such as
$\alpha^{27}$, but not easily all needed $O(p^4)$ LECs.\footnote{
In the more complicated case of $\Delta I=\half$ decays,
also total-derivative operators, such as
$\partial_\mu[\Sigma\partial_\mu\Sigma^\dagger,
\Sigma M^\dagger \pm M\Sigma^\dagger]_{\pm ds}$, appear at $O(p^4)$.  }  
The ``chiral logarithms" in eqs.~(\ref{phys},\ref{unphys}) are different
in the two cases, and, for typical lattice quark masses, they are
sizable.  This is an indication that one should at least use ChPT
to one loop in the analysis of lattice results for $K\to\pi\pi$
matrix elements.

To conclude this example, we note that there are also power-like
finite-volume corrections, here given for a spatial box $L^3$ with
periodic boundary conditions.  These corrections come from pion-%
rescattering diagrams \cite{glp}.  They may be large: for $f=160$~MeV,
$M_\pi=500$~MeV and $M_\pi L=6$ (typical lattice values of these parameters),
they are about $20\%$ of the tree-level contribution.
By comparing computations on different volumes, JLQCD found good
agreement between the prediction from ChPT of eq.~(\ref{unphys})
and results from lattice
QCD (first paper of \cite{num}).  
In general, one-loop ChPT gives a good fit of
the JLQCD results (which are quenched, at $a^{-1}=2$~GeV), with a 
reasonable value of $d_v$.  It appears that one-loop ChPT is sufficient
to explain the discrepancy \cite{bs} between unphysical 
$(K\to\pi\pi)_{\Delta I=3/2}$ lattice results 
converted to physical ones at tree level only of ChPT and the experiment.

A similar analysis can be carried out for $\Delta I=\half$
decays.  
This case is however more complicated for a variety of reasons
($O(p^4)$ LECs being only one of them -- for a nice review see 
ref.~\cite{cbtasi}), and it may be advantageous to use the
chiral-symmetry connection to $K\to\pi$ matrix elements.  At tree
level in ChPT, this works as follows \cite{cbetal2}.  The
desired $K\to\pi\pi$ matrix elements are given by
\begin{eqnarray}
\langle\pi^+\pi^-|27plet|K^0\rangle&=&\frac{4i}{f^3}(m_K^2-m_\pi^2)
\alpha^{27}\ ,\label{kpipi}\\
\langle\pi^+\pi^-|octet|K^0\rangle&=&-\frac{4i}{f^3}(m_K^2-m_\pi^2)
\alpha^8_1\ ,\nonumber
\end{eqnarray}
in terms of the $O(p^2)$ $(27_L,1_R)$ and $(8_L,1_R)$ LECs 
$\alpha^{27}$ and $\alpha^8_1$ respectively.  They
 can also be obtained from
\begin{eqnarray}
\langle\pi^+|27plet|K^+\rangle&=&-\frac{4}{f^2}M^2\alpha^{27}\ ,
\label{ktopi}\\
\langle\pi^+|octet|K^+\rangle&=&\frac{4}{f^2}M^2(\alpha^8_1-\alpha^8_2)\ ,
\nonumber
\end{eqnarray}
taking $M_K=M_\pi=M$ on the lattice.
The new (unphysical) LEC $\alpha^8_2$ can be determined from the matrix element
\begin{equation}
\langle 0|octet|K^0\rangle=\frac{4i}{f}(M_K^2-M_\pi^2)\alpha^8_2\ .
\label{ktovac}
\end{equation}
To lowest order in ChPT, $K\to\pi\pi$ matrix elements
can be therefore obtained from $K\to\pi$ and $K\to 0$, but the likely size
of chiral corrections on the lattice (and in nature) makes it
imperative to extend the analysis to, at least, next-to-leading order in ChPT.
At one loop, ChPT predicts a behavior like \cite{gp}
\begin{equation}
\langle\pi^+|octet|K^+\rangle=\frac{4M^2}{f^2}\left[
\alpha^8_1(1+{\rm logs})-\alpha^8_2(1+{\rm logs})
+A_v M^2+A_s NM_{sea}^2\right]\ ,\label{ktopiol}
\end{equation}
and analogous expressions 
for the other matrix elements.  Indeed, for typical lattice
masses, the logarithms can be large.  Hence, they and the polynomial
terms with coefficients $A_{v,s}$, which are linear combinations of
$O(p^4)$ LECs, will have to be taken into account in fits to lattice data.
Unfortunately, the combinations of $O(p^4)$ LECs contained in $K\to\pi$
matrix elements do not carry enough information to fully determine physical
$K\to\pi\pi$ matrix elements up to $O(p^4)$.  Also note that for
$K\to 0$ and $K\to\pi$ with $M_K\ne M_\pi$ total-derivative
operators can (and do) contribute.  In this context we mention that there 
also exist subtleties with  penguin operators in (partially) quenched QCD 
which have been discussed in ref.~\cite{gpp}.

One lesson to be learned is that, in our view, 
it should be possible to extract reliable values of the
$O(p^2)$ LECs $\alpha^8_1$ and $\alpha^{27}$ from the lattice computation of 
$K\to\pi$ transition matrix elements.
This is interesting in its own right, specially since phenomenological
estimates of these LECs already exist \cite{kmw}.  
In addition, we believe that the analysis of lattice results within the 
approaches considered here requires the
use of next-to-leading order ({\it i.e. $O(p^4)$}) ChPT for a variety
of reasons.  The most important of these are to check for convergence
of the chiral expansion, to reduce systematic errors coming from the
use of ChPT, and to understand finite-volume effects.


\begin{thebibliography}{99}
\bibitem{mg}
M.~Golterman,
hep-ph/0011084.
  
\bibitem{cbetal1}
C.~Bernard {\it et al.},
Nucl.\ Phys.\ Proc.\ Suppl.\  {\bf 4}, 483 (1988).

\bibitem{cbetal2}
C.~Bernard {\it et al.},
Phys.\ Rev.\ D {\bf 32}, 2343 (1985).

\bibitem{BPP}
 J.~Bijnens, E.~Pallante and J.~Prades,
 Nucl.\ Phys.\ B {\bf 521}, 305 (1998). 
 
\bibitem{num}
S.~Aoki {\it et al.}  [JLQCD Collaboration],
Phys.\ Rev.\ D {\bf 58}, 054503 (1998);
R.~Gupta, T.~Bhattacharya and S.~Sharpe,
Phys.\ Rev.\ D {\bf 55}, 4036 (1997);
D.~Pekurovsky and G.~Kilcup,
hep-lat/9812019;
A.~Donini, V.~Gimenez, L.~Giusti and G.~Martinelli,
Phys.\ Lett.\ B {\bf 470}, 233 (1999);
T.~Blum  [RBC collaboration],
Nucl.\ Phys.\ Proc.\ Suppl.\  {\bf 94}, 291 (2001);
R.~D.~Mawhinney  [RBC Collaboration],
Nucl.\ Phys.\ Proc.\ Suppl.\  {\bf 94}, 315 (2001);
A.~Ali Khan {\it et al.}  [CP-PACS Collaboration],
Nucl.\ Phys.\ Proc.\ Suppl.\  {\bf 94}, 283 (2001);

\bibitem{gm}
G. Martinelli, these proceedings.

\bibitem{ll}
L.~Lellouch and M.~L\"uscher,
Commun.\ Math.\ Phys.\  {\bf 219}, 31 (2001).

\bibitem{mt}
L.~Maiani and M.~Testa,
Phys.\ Lett.\ B {\bf 245}, 585 (1990).

\bibitem{LMST}
 C.-J.D.~Lin, G.~Martinelli, C.T.~Sachrajda and M.~Testa,
hep-lat/0104006.

\bibitem{glp}
M.~Golterman and K.-C.~Leung,
Phys.\ Rev.\ D {\bf 56}, 2950 (1997);
Phys.\ Rev.\ D {\bf 57}, 5703 (1998);
Phys.\ Rev.\ D {\bf 58}, 097503 (1998);
E.~Pallante,
JHEP {\bf 9901}, 012 (1999);
M.~Golterman and E.~Pallante,
Nucl.\ Phys.\ Proc.\ Suppl.\  {\bf 83}, 250 (2000).

\bibitem{shsh}
S.~Sharpe and N.~Shoresh,
Phys.\ Rev.\ D {\bf 62}, 094503 (2000).

\bibitem{bs}
C.~Bernard and A.~Soni,
Nucl.\ Phys.\ Proc.\ Suppl.\  {\bf 17}, 495 (1990).

\bibitem{cbtasi}
C.~Bernard,
in {\it ``From Actions to Answers,"} proceedings of TASI '89,
eds. T.~DeGrand and D.~Toussaint (World Scientific, 1990).

\bibitem{gp}
M.~Golterman and E.~Pallante,
JHEP {\bf 0008}, 023 (2000).

\bibitem{gpp}
M. Golterman and E. Pallante, hep-lat/0108010.

\bibitem{kmw}
J.~Kambor, J.~Missimer and D.~Wyler,
Phys.\ Lett.\ B {\bf 261}, 496 (1991).

\end{thebibliography}
\end{document}